\documentclass[dvips]{article}
\usepackage{supertabular,lscape,epsfig}
\usepackage{amssymb}
\usepackage{amsmath}

\DeclareSymbolFont{ppa}{OT1}{ppl}{m}{it}
\DeclareMathSymbol{\vv}{\mathalpha}{ppa}{'166}

\begin{document}

\newcommand{\dd}{\,{\rm d}}
\newcommand{\ie}{{\it i.e.},\,}
\newcommand{\etal}{{\it et al.\ }}
\newcommand{\eg}{{\it e.g.},\,}
\newcommand{\cf}{{\it cf.\ }}
\newcommand{\vs}{{\it vs.\ }}
\newcommand{\zdot}{\makebox[0pt][l]{.}}
\newcommand{\up}[1]{\ifmmode^{\rm #1}\else$^{\rm #1}$\fi}
\newcommand{\dn}[1]{\ifmmode_{\rm #1}\else$_{\rm #1}$\fi}
\newcommand{\upd}{\up{d}}
\newcommand{\uph}{\up{h}}
\newcommand{\upm}{\up{m}}
\newcommand{\ups}{\up{s}}
\newcommand{\arcd}{\ifmmode^{\circ}\else$^{\circ}$\fi}
\newcommand{\arcm}{\ifmmode{'}\else$'$\fi}
\newcommand{\arcs}{\ifmmode{''}\else$''$\fi}
\newcommand{\MS}{{\rm M}\ifmmode_{\odot}\else$_{\odot}$\fi}
\newcommand{\RS}{{\rm R}\ifmmode_{\odot}\else$_{\odot}$\fi}
\newcommand{\LS}{{\rm L}\ifmmode_{\odot}\else$_{\odot}$\fi}

\newcommand{\Abstract}[2]{{\footnotesize\begin{center}ABSTRACT\end{center}
\vspace{1mm}\par#1\par
\noindent
{~}{\it #2}}}

\newcommand{\TabCap}[2]{\begin{center}\parbox[t]{#1}{\begin{center}
  \small {\spaceskip 2pt plus 1pt minus 1pt T a b l e}
  \refstepcounter{table}\thetable \\[2mm]
  \footnotesize #2 \end{center}}\end{center}}

\newcommand{\TableSep}[2]{\begin{table}[p]\vspace{#1}
\TabCap{#2}\end{table}}

\newcommand{\FigCap}[1]{\footnotesize\par\noindent Fig.\  %
  \refstepcounter{figure}\thefigure. #1\par}

\newcommand{\TableFont}{\footnotesize}
\newcommand{\TableFontIt}{\ttit}
\newcommand{\SetTableFont}[1]{\renewcommand{\TableFont}{#1}}

\newcommand{\MakeTable}[4]{\begin{table}[htb]\TabCap{#2}{#3}
  \begin{center} \TableFont \begin{tabular}{#1} #4 
  \end{tabular}\end{center}\end{table}}

\newcommand{\MakeTableSep}[4]{\begin{table}[p]\TabCap{#2}{#3}
  \begin{center} \TableFont \begin{tabular}{#1} #4 
  \end{tabular}\end{center}\end{table}}

\newenvironment{references}%
{
\footnotesize \frenchspacing
\renewcommand{\thesection}{}
\renewcommand{\in}{{\rm in }}
\renewcommand{\AA}{Astron.\ Astrophys.}
\newcommand{\AAS}{Astron.~Astrophys.~Suppl.~Ser.}
\newcommand{\ApJ}{Astrophys.\ J.}
\newcommand{\ApJS}{Astrophys.\ J.~Suppl.~Ser.}
\newcommand{\ApJL}{Astrophys.\ J.~Letters}
\newcommand{\AJ}{Astron.\ J.}
\newcommand{\IBVS}{IBVS}
\newcommand{\PASP}{P.A.S.P.}
\newcommand{\Acta}{Acta Astron.}
\newcommand{\MNRAS}{MNRAS}
\renewcommand{\and}{{\rm and }}
\section{{\rm REFERENCES}}
\sloppy \hyphenpenalty10000
\begin{list}{}{\leftmargin1cm\listparindent-1cm
\itemindent\listparindent\parsep0pt\itemsep0pt}}%
{\end{list}\vspace{2mm}}

\def\TYLDA{~}
\newlength{\DW}
\settowidth{\DW}{0}
\newcommand{\dw}{\hspace{\DW}}

\newcommand{\refitem}[5]{\item[]{#1} #2%
\def\REFARG{#3}\ifx\REFARG\TYLDA\else, {\it#3}\fi
\def\REFARG{#4}\ifx\REFARG\TYLDA\else, {\bf#4}\fi
\def\REFARG{#5}\ifx\REFARG\TYLDA\else, {#5}\fi.}

\newcommand{\Section}[1]{\section{#1}}
\newcommand{\Subsection}[1]{\subsection{#1}}
\newcommand{\Acknow}[1]{\par\vspace{5mm}{\bf Acknowledgments.} #1}
\pagestyle{myheadings}

\def\thefootnote{\fnsymbol{footnote}}
%%%%%%%%%%%%%%%%%%%%%%%%%%%%%%%%%%%%%%%%%%%%%%%%%%%%%%%%%%%%%%%%%%%%%%%%%%%%

\begin{center}
{\Large\bf Variable Yellow and Red Stragglers in the 
Old Open Cluster NGC~6791}
\vskip1cm
{\bf
Janusz~~K~a~l~u~z~n~y}
\vskip3mm
{Copernicus Astronomical Center, Bartycka 18, 00-716 Warszawa, PL\\
e-mail: jka@camk.edu.pl}
\end{center}

\Abstract{Two sets of archive  time series observations
of an old open cluster NGC~6791 were re-analysed using
an image subtraction technique. We report identification of four new
variables. Proper motion data are available  for three of 
them and indicate that they are likely 
cluster members. Photometry of these stars is discussed along with
the data for two earlier identified variables. 
The sample analysed includes one yellow and three red stragglers.
An eclipsing red straggler V9 is of particular interest.
Cluster membership of this RS Cvn type binary  would imply that
its cooler component is in a stage of thermal inequilibrium.
One of variables is a K type red giant showing low-amplitude variability
with a period of about 0.33~d.  
}{~}

\Section{Introduction}%1
\vspace*{13pt}
With an age estimated at 8 Gyr (Chaboyer et al. 1999) a mass exceeding 
4000~m$_{\odot}$ (Kaluzny \& Udalski 1992) and metallicity 
$[{\rm Fe/H}]=0.32$ (Worthey \& Jowett 2003) the stellar 
cluster NGC~6791 may be considered as an intermediate object between 
old open clusters and globular clusters from the bulge region of the Galaxy.
The cluster hosts two cataclysmic variables (Kaluzny et al. 1997) and
is known to possess a bimodal horizontal branch with about 1/3 of
HB stars located on the blue horizontal branch (Kaluzny \& Udalski 1992;
Liebert, Saffer \& Green 1994).

There are a total of 74 known variable stars in the
field of NGC 6791. They were identified by Kaluzny and Rucinski 
(1993; hereafter KR),
Rucinski Kaluzny \& Hilditch (1996; hereafter RKH) and by  
Mochejska et al. (2002; 2003).
In particular, the  paper by Mochejska et al. (2002) reports first 
results of a large survey aimed at detection
of possible planetary transients in the cluster stars. The field of
massive search for planetary transients started with a 
survey of 47 Tuc conducted by Gilliliand et al. (2000) using the HST.
Recently the OGLE group reported detection of 121 stars with 
transiting low surface brightness companions (Udalski et al. 2002; 
Udalski et al 2002a; Udalski et al 2002b). 
So far one of these companions was confirmed 
to be a  body of planetary mass (Konacki et al. 2003). 
  
In this paper we present re-analysis of the data already used in 
KR and RKH. New time series photometry was extracted 
using an image subtraction technique. 
The main motivation for undertaking that job was an attempt to
detect some possible  planetary transients. It was also expected that a new 
precise photometry may lead to 
detection of some new variables in the cluster field.

\Section{Observing material and data reductions}%2

The first data set used in this contribution was collected with the 1-m
Jacobus Kapteyn Telescope during 14 consecutive nights 
in July 1995. Most of the data were taken during bright time. 
The instrument was a $1024 \times 1024$ CCD camera giving a scale of 
0.331 arcsec per pixel. A total of 619 and 154 frames were collected 
in the $V$ and $I$ bands, respectively. The exposure time was set to
5 min for all images. In the present analysis we used 579 and 139 best seeing
frames for $V$ and $I$ filters, respectively. Median seeing  for that 
subset of data was 1.16 arcsec for the $V$ band and  
1.09 arcsec for the $I$ band. 

The second data set was obtained on four consecutive nights in 
October 1991 using 0.9-m telescope at the Kitt Peak National Observatory.   
The instrument was a $1024 \times 1024$ CCD camera giving a scale of
0.68 arcsec per pixel.
The cluster was monitored for a total of about 16 hours. Present 
analysis is based on 61 and 8 images for the  $V$ and $B$ 
bands, respectively. The median seeing for the $V$ band  was
1.77 arcsec and the  exposure time was equal to 10 min for all frames. 
Exposures in the $B$ band lasted 15 min.

More details on both  data sets used in this paper 
are given in RKH and KR. 

Photometry was extracted using the image subtraction package 
ISIS V2.1 (Alard \& Lupton 1998; Alard 2000).
For each of the 4 filter and data-set combinations a reference image 
was constructed by combining several  individual best seeing images.
Daophot/Allstar package (Stetson 1987) was then used to 
get instrumental profile photometry for stellar objects detectable 
on reference images. 
Subsequently differential light curves expressed in ADU were derived with the 
ISIS routine $phot$ for all these stars.  
Finally, the light curves  were converted from ADU to the instrumental 
magnitudes defined by the  profile photometry extracted from the 
reference images. At that point we used the procedure which is described 
in detail in Mochejska et al. (2002).
% We would only like to add that 
%at this point appropriate aperture corrections need to be added to 
%the instrumental magnitudes derived with Daophot/Allstar. 

Transformation to 
the standard BVI system was performed based on a set of local standards
defined in the cluster field by Stetson (2000). 
For the KR data set usage of a total of 592 local standards resulted in the 
following transformations:
\begin{eqnarray}
v=V - 0.0021(B-V) - 3.0557\\
b=B - 0.1336(B-V) - 2.7664\\
b-v=0.2815+0.8763(B-V)
\end{eqnarray}
For the RKH data set usage of 399 local standards lead to relations:
\begin{eqnarray}
v=V-0.0553(V-I)-3.3728\\
i=I-0.0127(V-I)-2.8809\\
v-i=0.9577(V-I)-0.4922
   \end{eqnarray}

Time series photometry obtained for the $V$ band was the subject of analysis 
aimed at detection  of variable objects. First all light curves were 
searched for presence of periodic signal using the TATRY program 
provided kindly by Dr. Schwarzenberg-Czerny. That program makes
use of a multiharmonic periodogram method  which is described 
in Schwarzenberg-Czerny (1996). 
It is capable of finding different types of periodic variables
but is most suitable for detection of objects showing continuous 
changes of luminosity.
To search for possible eclipsing binaries
we used the $BLS$ method introduced recently by Kov\'acs et al. (2002).
It is particularly suitable for detection of 
shallow and/or narrow eclipses which can be potentially due to
planetary transients.

We have identified a total of 4 new variables in the RKH data set.
One of them, V76, was identified independently in the KR data set.
All variables reported in RKH and KR were recovered. 
We have failed to detect any candidates for planetary transients.
Equatorial coordinates of new variables are listed in Table 1. 
The transformation from rectangular to equatorial coordinates 
was derived using 375 stars from the USNO A-2 catalog (Monet et al. 1998) 
which were located in the $V$-band reference image for RHK data set.
We present finder charts for the newly discovered variables  in Fig. 1.

In Fig. 2 we present a plot of the standard deviation $\sigma $ 
versus the average
$V$ magnitude for all light curves extracted from the RKH data
set.  One may note that for the brightest stars
the achievable precision is about  0.004 mag and the $\sigma$ 
reaches 0.01 mag at $V\approx 17.2$. 

\begin{figure}[htb]
\centerline{\includegraphics[width=12cm]{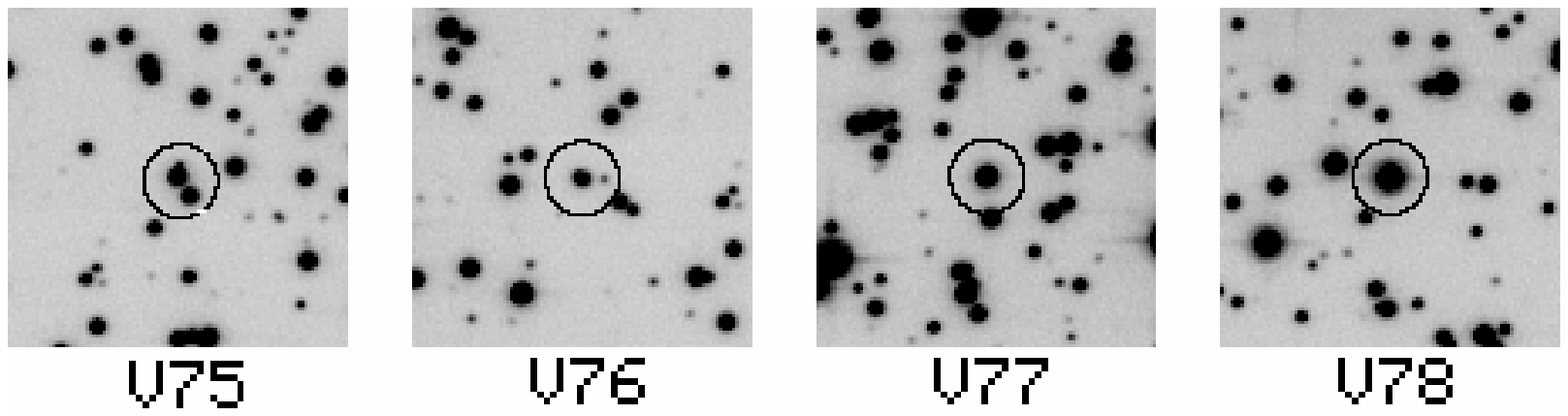}}
\vskip3mm
\FigCap{
Finder charts for the four newly detected variables. Each chart is 
33 arcsec on a side, with east to the left and north up.
}
\end{figure}

\begin{figure}[htb]
\centerline{
\includegraphics[width=10cm]{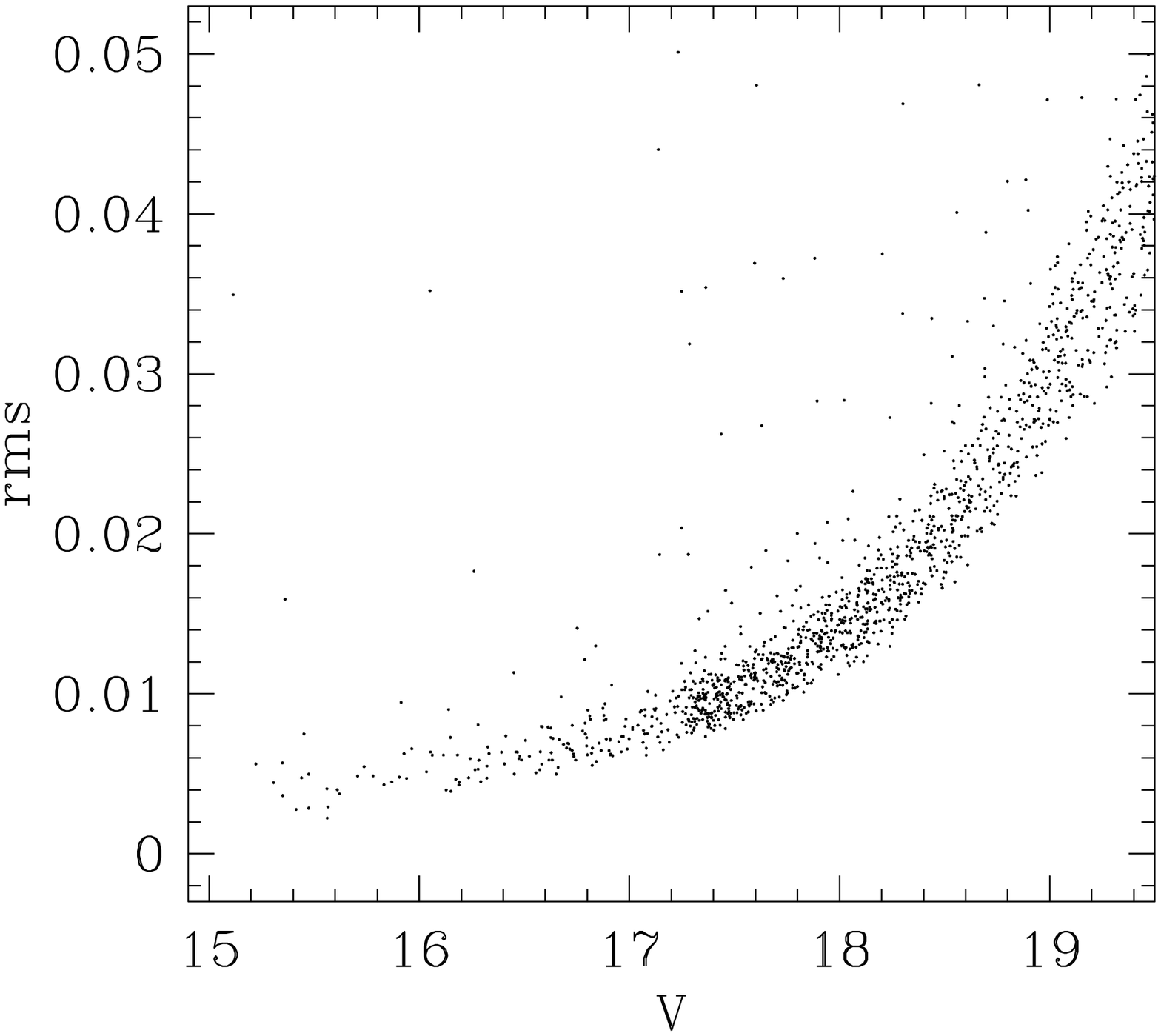}}
\vskip3mm
\FigCap{
Standard deviations of the light curves versus the average $V$ 
magnitudes for stars in the field of NGC~6791 (RKH data set).
}
\end{figure}

\MakeTable{lcc}{12.5cm}{Equatorial coordinates of newly identified binaries}
{
\hline
\noalign{\vskip2pt}
\multicolumn{1}{c}{ID} &  RA (J2000) [deg]& DEC (J2000) [deg]\\
\hline
\noalign{\vskip3pt}
V75 & 290.19982 & 37.76635  \\
V76 & 290.20778 & 37.76416   \\
V77 & 290.22046 & 37.77693   \\
V78 & 290.24108 & 37.78527   \\
\hline}

\Section{Properties of variables}%3

In Table 2 we provide some information about basic photometric properties
of four newly identified variables. 
The sample is supplemented with the data for two earlier identified
objects, V9 and V17, which will be  discussed below. 
Columns 2, 3 and 4 list maximal $V$ magnitude and average 
colors. The full range of observed luminosity changes  
is listed in columns 5 and 6 for the $V$ and $I$ band, respectively. 
Derived periods along with their
errors are listed in column 7. The last column of Table 2 gives 
cluster membership probabilities extracted from an unpublished
proper motion study conducted by Dr. Kyle Cudworth. 
Five  variables are likely members of NGC 6791 while 
for one of them, V75, proper motion data are not available.

\MakeTable{llllllll}{12.5cm}{Photometric data and membership probability for 
NGC~6791 variables }
{
\hline
\noalign{\vskip2pt}
\multicolumn{1}{c}{ID} &  $V_{\rm max}$& $<V-I>$& $<B-V>$& $\Delta V$ & $\Delta I$ &P [d] & MP[\%]\\
\hline
\noalign{\vskip3pt}
V9  & 17.25  & 1.40  &1.22 &0.23  &0.16   &3.187085(2)& 82\\ 
V17 & 17.915 & 1.267 &1.15 &0.091 &0.060  &6.20(5)        & 88 \\ 
V75 & 17.374 & 0.987 &0.92 &0.032 &0.017  &11.34(15)       & -\\
V76 & 18.226 & 1.140 &1.02 &0.094 &0.066  &4.19(2)       & 97\\
V77 & 16.674 & 0.953 &0.89 &0.014 &0.014  &7.7(1)        & 80\\
V78 & 15.526 & 1.354 &1.33 &0.009 & -     &0.332(3)     & 94 \\
\hline}

Figure 3 shows location of all variables in the $V/V-I$ and $V/B-V$
color magnitude diagrams of the cluster field. Positions marked correspond
to $V_{\rm max}$ and average color of variables.

Time domain light curves of V17 and V75-77 are 
shown in Fig. 4. Phased light curve of V78
is presented in Fig. 5. All light curves are based on 
the RKH data set.

In Fig. 6 we 
present phased light curves of V9 obtained
in 1991 (KR data set), 1995 (RKH data set) and 
2001 (Mochejska et al. 2002) seasons. Phased color curve of V9
from  the 1995 season  is shown in Fig. 7.

V75 is located slightly to the red of the cluster main sequence tip
on the color-magnitude diagram. Its 1995 light curve shows 
sine-like modulation with $P=11.34$~d and $17.374<V<17.406$. 
The data from the 1991 season (KR data set) 
give the  light curve with nightly averaged magnitudes from the range
$17.343<V<17.351$. The stars may belong to ellipsoidal variables
which would imply an orbital period of about 22.7 days. 
Presence of ellipsoidal variability in case of such long orbital period 
is conceivable only if V75 is composed of a slightly evolved star and
a dark companion with a mass exceeding significantly mass of the cluster
turnoff stars.\footnote{For the binary composed of two turnoff stars
and with $P_{\rm orb}\approx23$~d ellipsoidal variability would be 
much smaller than it is observed.}
Another possibility is that  V75 belongs to diverse class of 
"spotted variables". 

V78 is  located on the red giant branch
about 2 magnitudes above the turn-off region of the cluster. 
The proper motion  data indicate cluster membership with 
propbability of 94\%. The derived period
$P=0.33$~d is an alias of a 1~day period and the power spectrum 
of the light curve has a prominent peak also at $P\approx 0.25$~d. 
Unfortunately, in the KR  data 
set  the star was located close to a bad column on most images which
prevented derivation of its light curve. We also did not manage
to extract the $I$ band light curve from the 1995 data  as 
the variable was overexposed on all images in that filter. Images in
the $V$-band  were free from that problem. The value of the $V-I$ color 
listed for the variable in Table 2 was adopted from Stetson (2000).\\
Photometric variability at the 0.01 mag level is quite common among 
field giants of spectral type G-K. Henry et al. (2000) detected 
variability on time scales from 2 days to a few week weeks in a total of 81 
objects from a sample including 187 stars. Observed changes of luminosity 
were explained by presence of radial pulsations. \\
Edmonds and Gilliland (1996) detected 15 K giant variables
in the globular cluster 47 Tuc. Observed changes
had amplitudes 0.005-0.015 mag in the $U$ band and periods between
about 2 and 4 day. As a likely cause of variability the authors considered
radial or nonradial pulsations.\\
However, variability on a time scale shorter that 1 day was reported 
so far for a very few giants. Hatzes \& Cohran(1994) reported $\beta$~Oph 
to exhibit multiperiodic radial velocity variations with a period
of about 0.25~day. 
Observed amplitudes of radial velocity
correspond to predicted  photometric variablity at the level of a 
few mmag at best. Similarly, radial velocity variations with 
periods of the order of hours detected in $\eta$~Boo (Kjeldsen et al. 
2002) and in $\xi$~Hya (Frandsen et al. 2002) imply photometric effects
at the $\mu$ magnitude level. Both mentioned giants show supposedly
solar-like oscillations. \\   
In that context V78 may prove to be very a interesting object but 
follow-up observations are needed to confirm detected variability 
and to establish its period with confidence. 

V77 is a likely yellow straggler member of the cluster. The 
light curve shows sine-like modulation with the period
of about 7.7~d and  low amplitude. Variability of V77 
can be explained by an ellipsoidal effect in a binary system
or alternatively by "spots" related activity.

\subsection{Variable red stragglers}

Variables V9, V17 and V76 are all likely cluster members and moreover
they can be classified as  red stragglers. The  term "red straggler" 
is used to denote objects located to the red of
the main sequence or  subgiant branch on color-magnitude 
diagrams of old stellar clusters. They should not be confused 
with ordinary binary members of clusters which are also observed to 
the red of main sequences. Red stragglers may be defined
as objects whose position in the red part of the  color magnitude 
diagram of a parent cluster cannot be reproduced by combining 
light of two  ordinary stars from principal sequences.
Variable red stragglers were identified in 47 Tuc by Albrow 
et al. (2001) and more recently by Mathieu et al. (2003) in M67. 
Both red stragglers in M67 are binary stars.
Also an optical counterpart of the binary millisecond pulsar
PSR J1740-5340 in NGC 6397 (Ferraro et al. 2001) is an example of a  
red straggler. 
  
V9 is the only red straggler from NGC 6791 which is  certainly 
a binary star. Its  light curves from 1991 (KR data set), 1995 
(RKH data set) and 2001 (Mochejska et al. 2002) seasons are shown in Fig. 6.
The color curve obtained in 1995 is presented in Fig. 7.
The  light curves  were phased
using an ephemeris:\\
$Min I = HJD 244 8540.5739\pm 0.0013 + E\times 3.1870848\pm 0.0000016~d$\\
which is based on moments of 3 individual primary minima.
These minima span the period from 1991 to 2001 and are listed 
in Table 3. 

\MakeTable{lll}{12.5cm}{Moments of primary minima of V9}
{
\hline
\noalign{\vskip2pt}
\multicolumn{1}{c}{T[HJD]} &  $\sigma[d]$& E\\
\hline
\noalign{\vskip3pt}
244 8540.60 & 0.01 & 0 \\
244 9920.571 & 0.001 & 433  \\
245 2106.9259& 0.0003 & 1119  \\
\hline}

The 1995 and 2001 light curves of V9 phased with the new ephemeris
clearly show the presence of a shallow secondary eclipse centered at 
phase $\phi=0.50$. They also exhibit  significant asymmetry. The maximum 
following the primary eclipse is higher that that following the seacondary 
eclipse.
The disturbed shape of the light curve precludes determination of geometrical
elements of the binary by means of a standard light curve solution.
The color curve from the 1995 season shows noticeable dips during 
both eclipses. This is quite an  unusual behavior as in most eclipsing
binaries the observed color becomes bluer at the secondary 
eclipse.  At the primary eclipse the observed color of V9 is $V-I\approx 1.47$.
This value sets a lower limit on the color of the cooler component
of the system and places it at least  0.25 mag to the red of the 
cluster subgiant branch (see Fig. 3). 
If the binary is indeed a member of the cluster then
we may infer that its secondary is in stage of thermal inequilibrium.
%Such un-equilibrium could be due to present of recent occurence 
%of vigorous mass transfer in the system. 
%Spectroscopic observations and search 
%for changes of the orbital period
%of the binary would be  interesting in that context.

Large and variable asymmetry of the light curve of V9 (see Fig. 6) 
is most likely due to occurrence of starspots on the surface of 
the cooler component of the binary. Profound spot activity is very 
often observed in close binaries composed of  red subgiants (or giants)  and 
main sequence stars. That class of variable stars in known as RS CVn 
type binaries. Light curves of RS CVn binaries occasionally show 
disturbances reaching a few tenths of  magnitude in the $V$ band 
(Padmakar \& Pandey 1999).

%We speculate that V9 is about to  
%enter  the common envelope phase of the binary evolution
%(eg. Beer and Davies 2003).
%Spectroscopic observations and search for changes of the orbital period 
%of the binary would be  interesting in than context.
 
V17 shows light variations with period $P=6.20$~d. 
The 1995 season light curve is slightly asymmetric with some  
kind of a "bump" occuring about 1 day before the minimum light. 
The full range of magnitudes 
diminishess from 0.091 in the $V$-band to 0.069 in the $I$-band. 
The light curve in the $R$-band obtained in 2001 by Mochejska et al. 
(2001) shows also sine-like shape but with different kind of
asymmetry than is observed in our data from the 1995 season. 
There is no indication of
any detectable change of period of variability between the 1995 and the 2001 
seasons. If the variable is indeed a member of the cluster then it 
is difficult to explain its position on the color-magnitude diagram
without postulating that it is a binary whose evolution was affected
by some mass transfer between components. We note that V17 resembles
S1113 in M67 (Mathieu et al. 2003) in respect of position 
on the color magnitude diagram of the parent cluster as well as  
a character of observed luminosity changes. To explain variability of S1113
Mathieu et al. (2003) postulated that they are due to combined effects 
of ellipsoidality and stellar spots. These two effects allow
to explain also characteristic of the light curves of V17 and in 
particular their asymmetry and  season-to-season changes of 
shape.

V76  shows light variations with the period $P=4.19$~d.
For the 1995 season the full range of magnitudes 
changes from 0.094 in the $V$-band to 0.066 mag in the $I$-band.
The variable is located above an apparent  sequence of 
photometric binaries 
visible on both presented color-magnitude diagrams. Its $V$-band 
light curve for the 1991 season (KR data set) is shown in Fig. 8.  
During that season the average magnitude of the variable was
$<V>\approx  18.24$ what can be compared with $<V>=18.27$ 
observed in  the 1995 season. We would like to note that 
the variable possesses close visual companion at distance
$d=0.56$~arcsec. This companion has $V=20.43$, $B-V=1.18$
and $V-I=1.415$.

\Section{Summary}%4

In this paper we have discussed photometry of  six variables
from the central field of the old open cluster NGC~6791.
All but one of them can be considered  very likely cluster members
based on their observed proper motions. The sample includes three
red stragglers. Their location on the cluster color-magnitude
diagram is hard to explain without postulating that they
are evolved close binaries. In fact, variable red straggler V9
is an eclipsing close binary. Hypotheses about the
binary nature of red stragglers V17 and V76 could be tested 
by obtaining  suitable spectrosopic observations.\\ 
For the red giant variable V78,  our data indicate modulation 
of the light curve with a period of about 0.33~day and an amplitude
of about 0.005 mag in the $V$ band. This interesting object
seems to be an attractive target for further  observations.
The yellow straggler V77 shows modulation of the light curve
with a period of about 7.7~days.

\Acknow{This work was supported with
NSF grant AST-9819787 and polish  KBN grant 5P03D004.21. 
We are very grateful to Dr. Kyle Cudworth for
providing us with the proper motion data for NGC 6791.
We would also like to thank an anonymous referee, for a prompt
and useful report.}

%\newpage
% REFERENCES

%%%%%%%%%%
\clearpage

\begin{figure}[h]
\center{
\includegraphics[height=217mm,width=153mm]{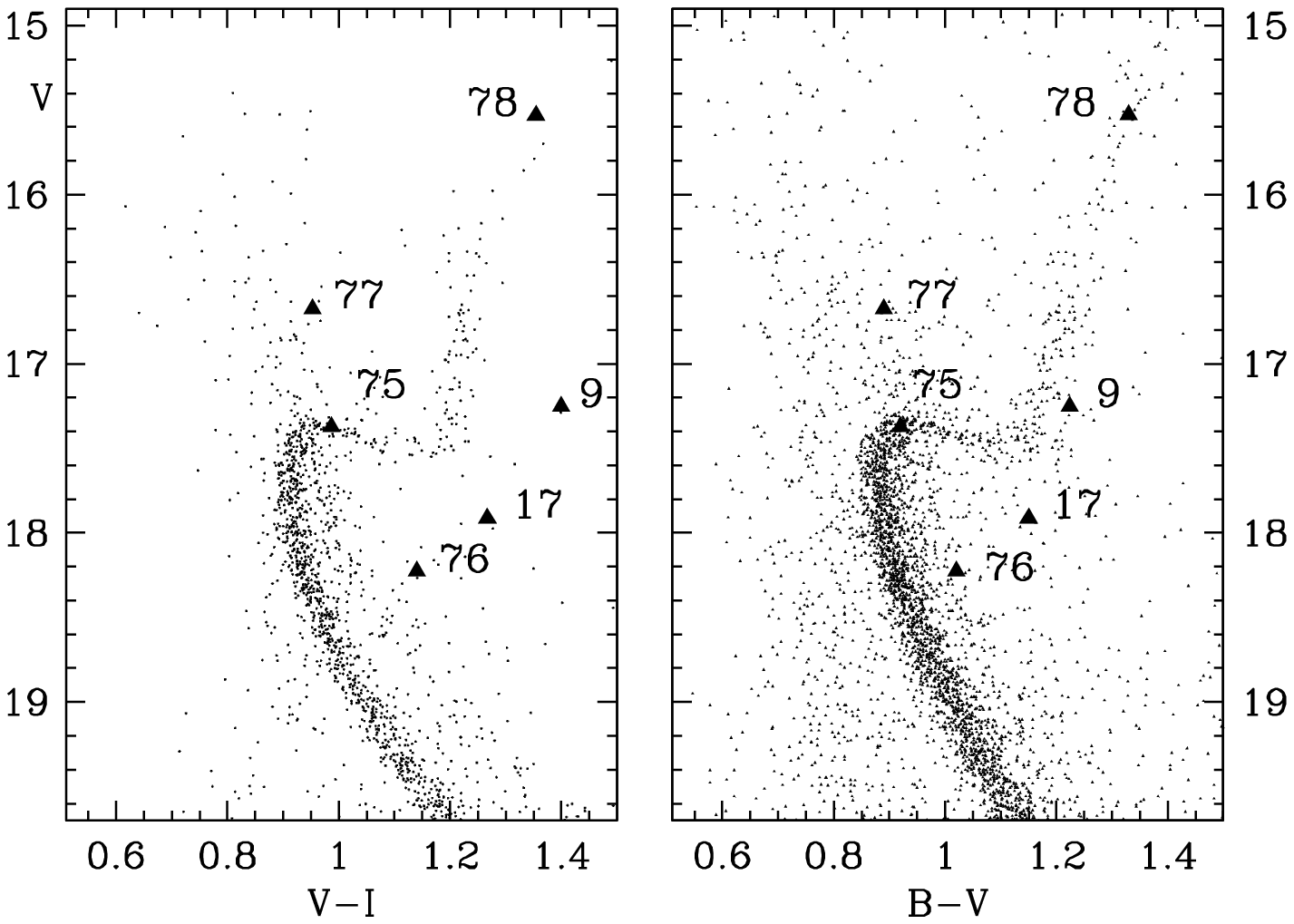}%
}
\caption{
The color-magnitude diagrams of NGC~6791 with positions
of the variables discussed in the paper marked. 
}
\end{figure}

\clearpage

\begin{figure}[h]
\center{
\includegraphics[height=216mm,width=173mm]{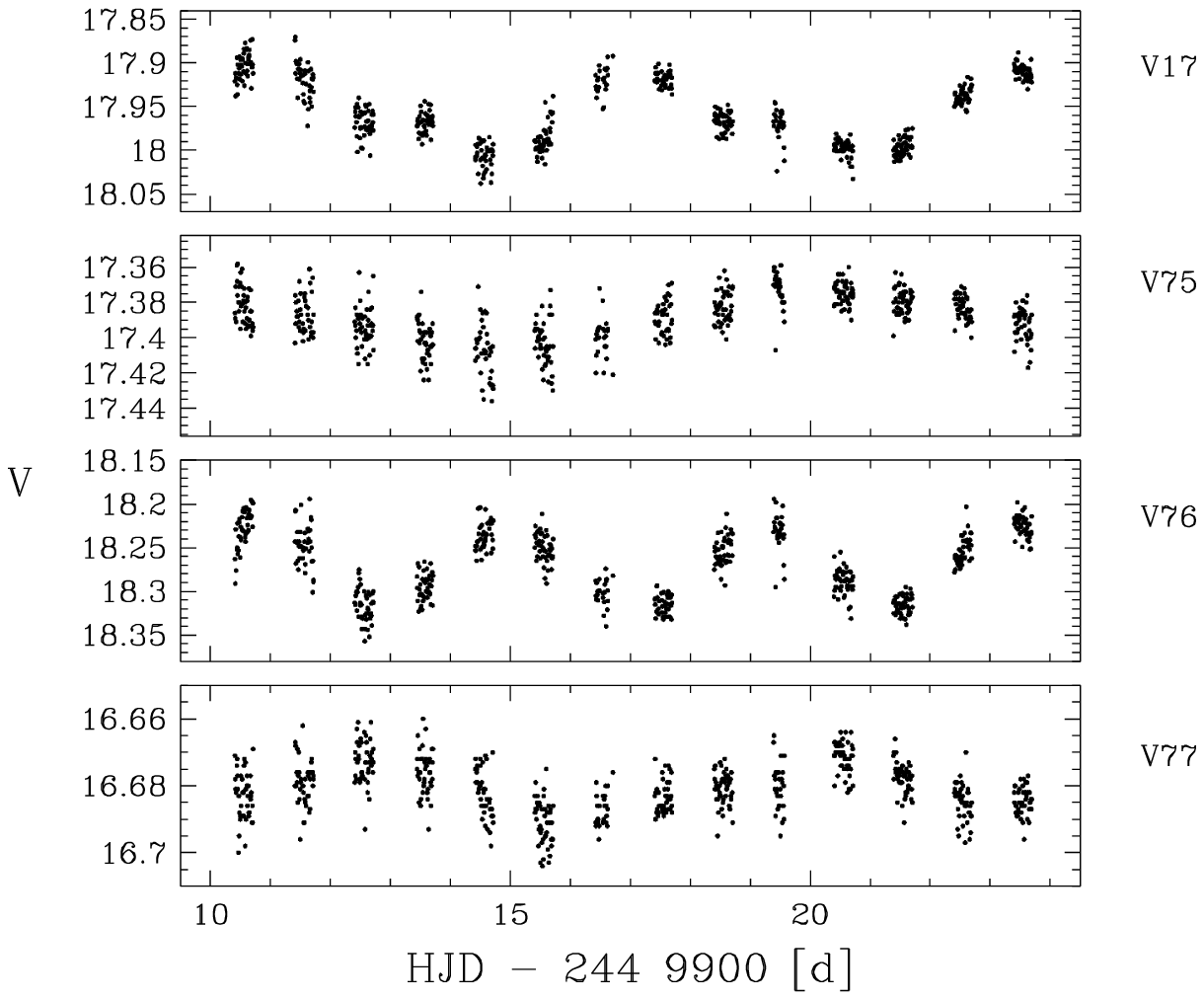}%
}
\caption{
The $V$ light curves of V17, V75, V76 and V77 based on the RKH data.
}
\end{figure}

\clearpage

\begin{figure}[h]
\center{
\includegraphics[height=216mm,width=173mm]{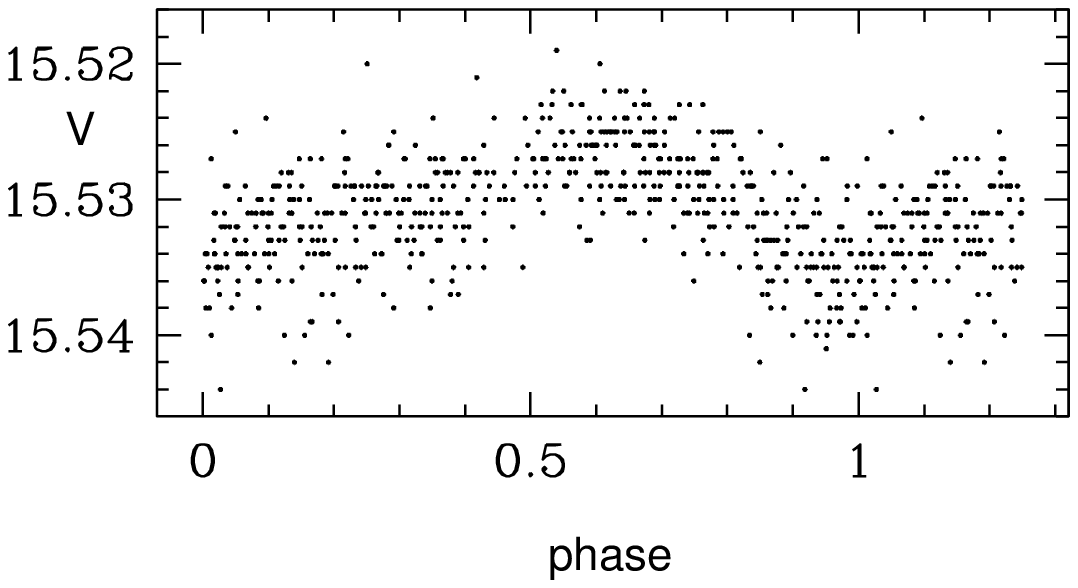}%
}
\caption{
The $V$ light curve of V78 for the 1995 season.
}
\end{figure}

\clearpage

\begin{figure}[h]
\center{
\includegraphics[height=216mm,width=173mm]{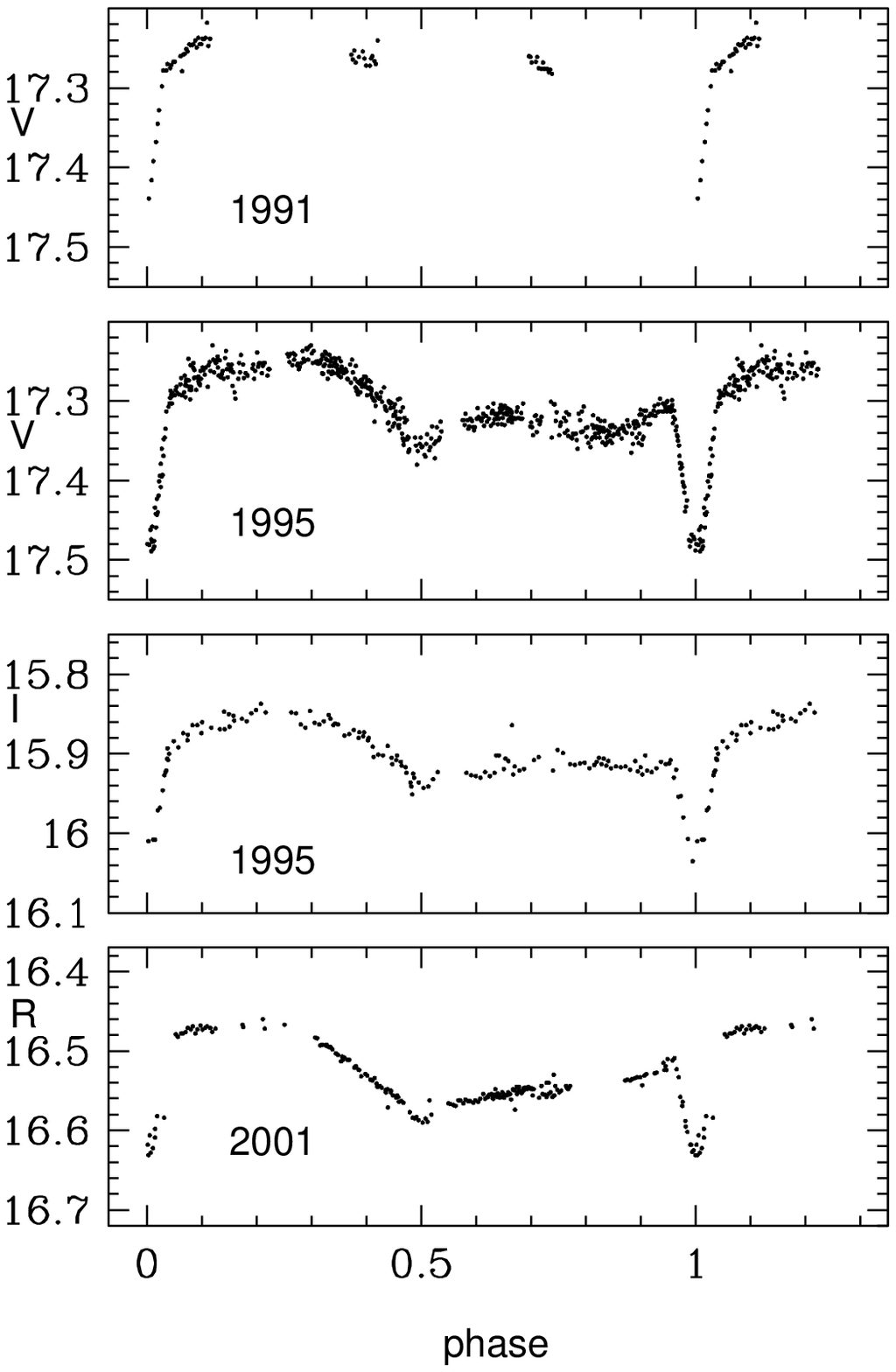}%
}
\caption{
Phased light curves of V9 for the observing seasons 1991, 1995
and 2001. Note the same scale for all panels.
}
\end{figure}
\clearpage

\begin{figure}[h]
\center{
\includegraphics[height=216mm,width=173mm]{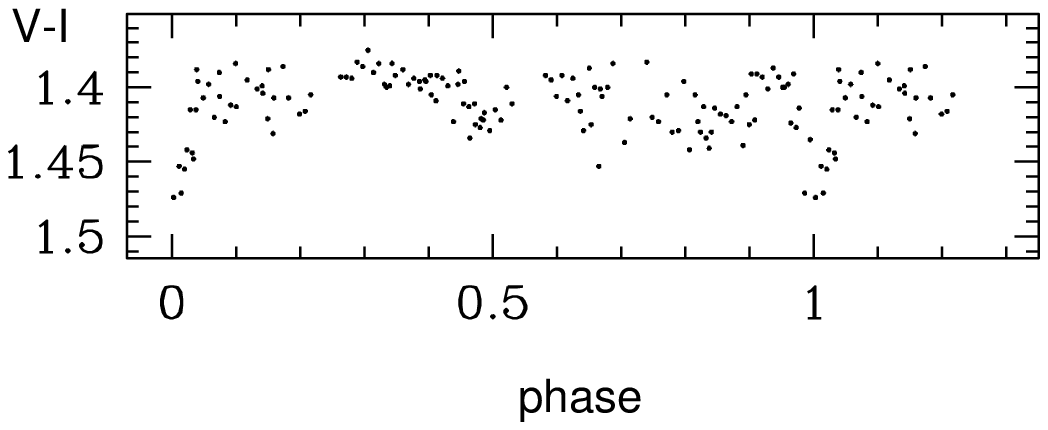}%
}
\caption{
Phased color curve of V9 from the observing seasons 1995.
}
\end{figure}

\clearpage

\begin{figure}[h]
\center{
\includegraphics[height=217mm,width=153mm]{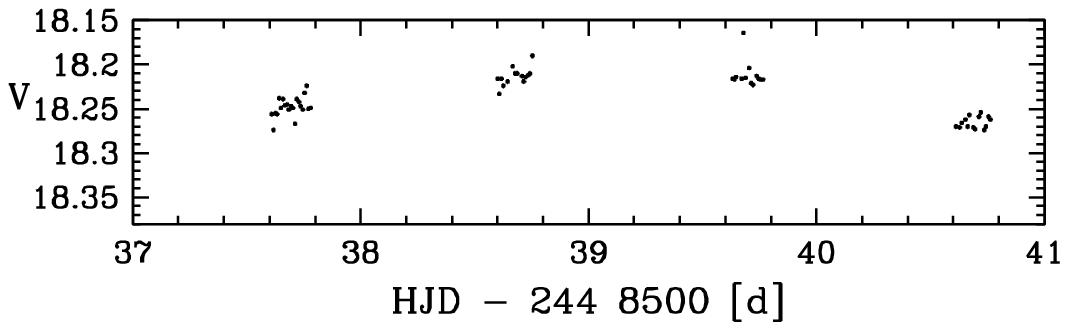}%
}
\caption{
The $V$ light curve of V76 for the 1991 season.
}
\end{figure}

\end{document}